\documentclass[11pt,a4paper]{article}

\usepackage[T1]{fontenc}
\usepackage[utf8]{inputenc}
\usepackage[margin=1in]{geometry}
\usepackage{amsmath,amssymb}
\usepackage{graphicx}
\usepackage{adjustbox}
\usepackage{xcolor}
\usepackage{url}
\usepackage[colorlinks=true,linkcolor=blue,citecolor=blue,urlcolor=blue]{hyperref}

\title{Hyperbolic embedding of higher-order simplicial networks}

\author{%
Martin Guillemaud$^{1,*}$, Vito Latora$^{2,3}$ and Mario Chavez$^{4}$\\[8pt]
\parbox{0.92\textwidth}{\centering\small $^{1}$Paris Brain Institute (ICM), CNRS UMR7225, Inserm U1127, Sorbonne University UM75, AP-HP. Piti\'e Salp\^etri\`ere University Hospital. Paris, France}\\[3pt]
\parbox{0.92\textwidth}{\centering\small $^{2}$School of Mathematical Sciences, Queen Mary University of London, London E1 4NS, United Kingdom}\\[3pt]
\parbox{0.92\textwidth}{\centering\small $^{3}$Dipartimento di Fisica ed Astronomia, Universit\`a di Catania and INFN, Catania I-95123, Italy}\\[3pt]
\parbox{0.92\textwidth}{\centering\small $^{4}$CNRS, Piti\'e Salp\^etri\`ere University Hospital. Paris, France}\\[8pt]
{\small $^{*}$Corresponding author: \texttt{martin.guillemaud@gmail.com}}}
\date{}

\begin{document}

\maketitle

\begin{abstract}
Many biological interactions involve groups rather than pairs, as when several proteins assemble into a multi-protein complex. Simplicial complexes naturally capture these collective interactions. 
Existing 
hyperbolic embeddings 
represent network hierarchy and similarity through a few interpretable coordinates,
but 
rely
only on pairwise links: they cannot distinguish three units engaged in a common interaction from three units 
connected only in pairs. 
We introduce here the hyperbolic higher-order (HHO) embedding, which embeds a simplicial complex into the Poincar\'e disk, 
while incorporating its group interactions. 
The hyperbolic area of the triangle spanned by three interacting nodes then measures how compactly their interaction is represented. 
Tests
on synthetic and real-world networks, including a co-authorship network and the protein interaction network of \textit{Escherichia coli}, show that HHO reduces the hyperbolic area of triangles representing simplices relative to those obtained from pairwise network embeddings. HHO also recovers node groups
more accurately  
than pairwise embeddings  and a weighted-clique alternative, with performance competitive with
Euclidean embeddings, while providing a low-dimensional representation that can be analyzed with the tools of network geometry.
\end{abstract}


\section{Introduction}\label{Sec:intro}
Complex networks provide a powerful framework to describe the structure of a wide range of real systems, from technological infrastructures to social, biological, and brain networks~\cite{Latora2017,Torres2021}. A central question, both from theoretical and applied perspectives, concerns the latent geometry that underlies the observed connectivity patterns. In this context, hyperbolic geometry has emerged as a particularly fruitful paradigm for modeling real complex networks, mainly because of its ability to simultaneously represent hierarchies, degree heterogeneity, and community structures~\cite{Krioukov2010,Bogu2010,Bogu2021,Garca-Prez2019}.

In parallel, the study of collective phenomena in complex systems has progressively moved beyond the framework of purely pairwise interactions to incorporate higher-order interactions, modeled by hypergraphs or simplicial complexes~\cite{Battiston2020,Battiston2021}. This growing recognition of the importance of higher-order interactions in complex systems highlights the need for novel analytical tools that jointly capture latent geometry and higher-order connectivity, and that can be applied across diverse domains, from synthetic networks to real interconnected systems. The present work contributes to this direction by introducing a general method to embed simplicial complexes in hyperbolic space through a multilayer representation.

The seminal work of Krioukov and collaborators showed that many complex networks can be understood as realizations of geometric graphs in hyperbolic space~\cite{Krioukov2010}: nodes have latent coordinates in a negatively curved space, and the probability of a link depends on a combination of popularity, encoded by the radial coordinate, and similarity, encoded by the angular coordinate~\cite{Papadopoulos2012}. This framework reproduces heavy-tailed degree distributions, strong clustering and hierarchical organization within a single geometric model~\cite{Krioukov2010,Bogu2021}, and the map of the Internet inferred from it accounts simultaneously for its topology and its routing properties~\cite{Bogu2010}. Network geometry has since been formalized as a general framework relating topology to latent geometry~\cite{Bogu2021}, and hyperbolic coordinates can be inferred from real network data, either by maximum likelihood, as in Mercator~\cite{Garca-Prez2019}, or by manifold learning, as in coalescent embeddings~\cite{Muscoloni2017}. Hyperbolic geometry has attracted considerable interest in machine learning as a framework for representing hierarchical and relational data. Notable examples include Poincaré embeddings~\cite{Nickel2017} and hyperbolic graph neural networks~\cite{Chami2019}. Its suitability for hierarchical representations is supported by the fact that trees can be embedded in hyperbolic space with arbitrarily low distortion, even in low dimensions~\cite{Sala2018}.

In many complex systems, interactions occur among groups of entities rather than exclusively between pairs: several authors co-sign a scientific article, multiple species compete for shared resources, and proteins interact collectively within molecular complexes, naturally giving rise to higher-order (HO) network representations based on hypergraphs and simplicial complexes. Hypergraphs and simplicial complexes provide the natural language to describe such higher-order interactions~\cite{Battiston2020,Battiston2021,Bick2023}. In a simplicial complex, every subset of a group is itself a group, which makes the tools of algebraic topology available~\cite{Bianconi2021}. Group interactions qualitatively change collective dynamics. In simplicial contagion models, in which transmission requires the joint presence of several neighbors, the transition to an endemic state becomes discontinuous, a behavior that pairwise models do not reproduce~\cite{Iacopini2019}; higher-order Kuramoto dynamics on simplicial complexes can make synchronization explosive~\cite{Millan2020}; and in models of ecological competition, interactions among more than two species stabilize the coexistence of large communities~\cite{Grilli2017}. Group interactions also have a structural signature. In co-authorship and many other datasets, new group interactions preferentially form among nodes that are already linked, a process of simplicial closure that makes higher-order structure partly predictable~\cite{Benson2018}, and persistent homology characterizes the higher-order organization of scientific collaborations across disciplines~\cite{Patania2017}. In neuroscience, for instance, cliques and cavities reveal aspects of the higher-order organization of brain networks that are invisible at the level of simple pairwise interactions~\cite{Giusti2016,Petri2014,Sizemore2018,Reimann2017}. In all these cases, the group interactions carry information that their pairwise representation does not.

Cellular networks offer a particularly natural setting in which latent geometry and higher-order structure coexist. Many biological processes involve more than two partners, such as a protein complex assembled from several subunits or a metabolic reaction involving multiple substrates and products, and are therefore not faithfully represented by pairwise edges\cite{Klamt2009}. At the same time, the metabolic networks of \textit{E.~coli} and human cells have been shown to be congruent with an underlying hidden geometry in which proximity encodes structural similarity\cite{Serrano2012}. Protein interaction networks, in which experimentally characterized complexes define higher-order groups on top of pairwise interactions, therefore constitute a natural testbed for geometric and higher-order methods.

Despite these parallel advances, the two lines of research (hyperbolic geometry and HO representations) have remained largely disconnected. Existing hyperbolic embedding methods for network data, whether based on maximum likelihood~\cite{Bogu2010,Garca-Prez2019}, on manifold learning~\cite{Muscoloni2017,AlanisLobato2016}, or on gradient-based optimization~\cite{Nickel2017}, take as input a graph of pairwise connections. Applying them to a simplicial complex therefore requires first reducing it to its 1-skeleton, the graph of its edges. This reduction is lossy: a filled triangle, which represents a genuine three-body interaction, and an empty triangle, formed by three independent pairwise links, have the same 1-skeleton and are thus embedded identically. More generally, higher-order interactions are often represented by projecting each group onto a clique in a pairwise network. However, this projection discards information about group membership and introduces pairwise links that may not correspond to actual dyadic interactions, thereby artificially increasing network density~\cite{Kitsak2017}. 

Standard reductions of hypergraphs to graphs, such as clique and star expansions or the normalized hypergraph Laplacian~\cite{Agarwal2006,Zhou2006}, can encode group membership, but they were developed for spectral clustering on Euclidean embedding spaces. Conversely, hyperbolic graph neural networks~\cite{Chami2019} have recently been extended to hypergraphs~\cite{Bai2025}, but these are supervised feature-learning models optimized for prediction tasks; they do not provide an interpretable latent geometry in which the position of each node and the footprint of each higher-order interaction can be read and quantified. Although generative models of growing simplicial complexes, in which new simplices are attached along existing faces, can produce networks with an underlying hyperbolic geometry~\cite{BianconiRahmede2017}, a general method for embedding an observed simplicial complex into hyperbolic space while preserving its higher-order organization and providing an interpretable characterization of its latent geometry remains, to the best of our knowledge, lacking.

Here we introduce such a method, the hyperbolic higher-order (HHO) embedding. Its central idea is to encode a simplicial complex as a two-layer network: the first layer is the graph of pairwise interactions, and the second layer contains one node per filled triangle, i.e.\ per triangle whose three nodes take part in a common group interaction, coupled to each of its three vertices and to the filled triangles with which it shares an edge. The resulting multilayer network is then embedded in the Poincar\'e disk with our recently introduced method for multilayer hyperbolic embedding~\cite{GuillemaudMultilayer2025}. Throughout the article, HHO denotes both the two-layer network representation and its embedding, and the pairwise embedding denotes the same hyperbolic embedding applied to the 1-skeleton alone (section~\ref{Sec:method_stats}). Accordingly, all comparisons between the two approaches assess the contribution of higher-order information to the resulting hyperbolic representation.

To quantify how higher-order interactions are represented in the latent space, we measure the hyperbolic area of the triangle spanned by the embedded nodes of each 2-simplex, which in curvature $-1$ equals its angular defect, $A = \pi - (\alpha_1+\alpha_2+\alpha_3)$, where $\alpha_1,\alpha_2,\alpha_3$ are its interior angles. A small area indicates that the embedding places jointly interacting units in a small compact region of the latent space. Comparing the areas of filled and empty triangles thus provides a direct, interpretable marker of whether higher-order organization has been captured, and a geometric counterpart to the compactness of functional modules and to the economy of wiring observed in biological systems~\cite{Bullmore2012}. Our approach leverages hyperbolic geometry to provide an interpretable representation of higher-order networks, in which their higher-order organization is explicitly encoded in the latent geometry and can be quantitatively characterized, complementing existing embedding approaches.

Our contributions are: (i) a transformation of simplicial complexes into two-layer networks and its embedding in hyperbolic space; (ii) surface-based markers comparing filled and empty triangles in the latent space; (iii) a validation on synthetic complexes with planted surfaces, on a co-authorship complex and on a case study of protein complexes in the protein interaction network of \textit{E.~coli}; and (iv) the open-source implementation of the method in the Python package HOHypEmb, together with example notebooks that compute the embeddings reported in this article.

\section{Method}\label{Sec:method}
\begin{figure}[!htbp]
    \centering
    \includegraphics[width=0.9\textwidth]{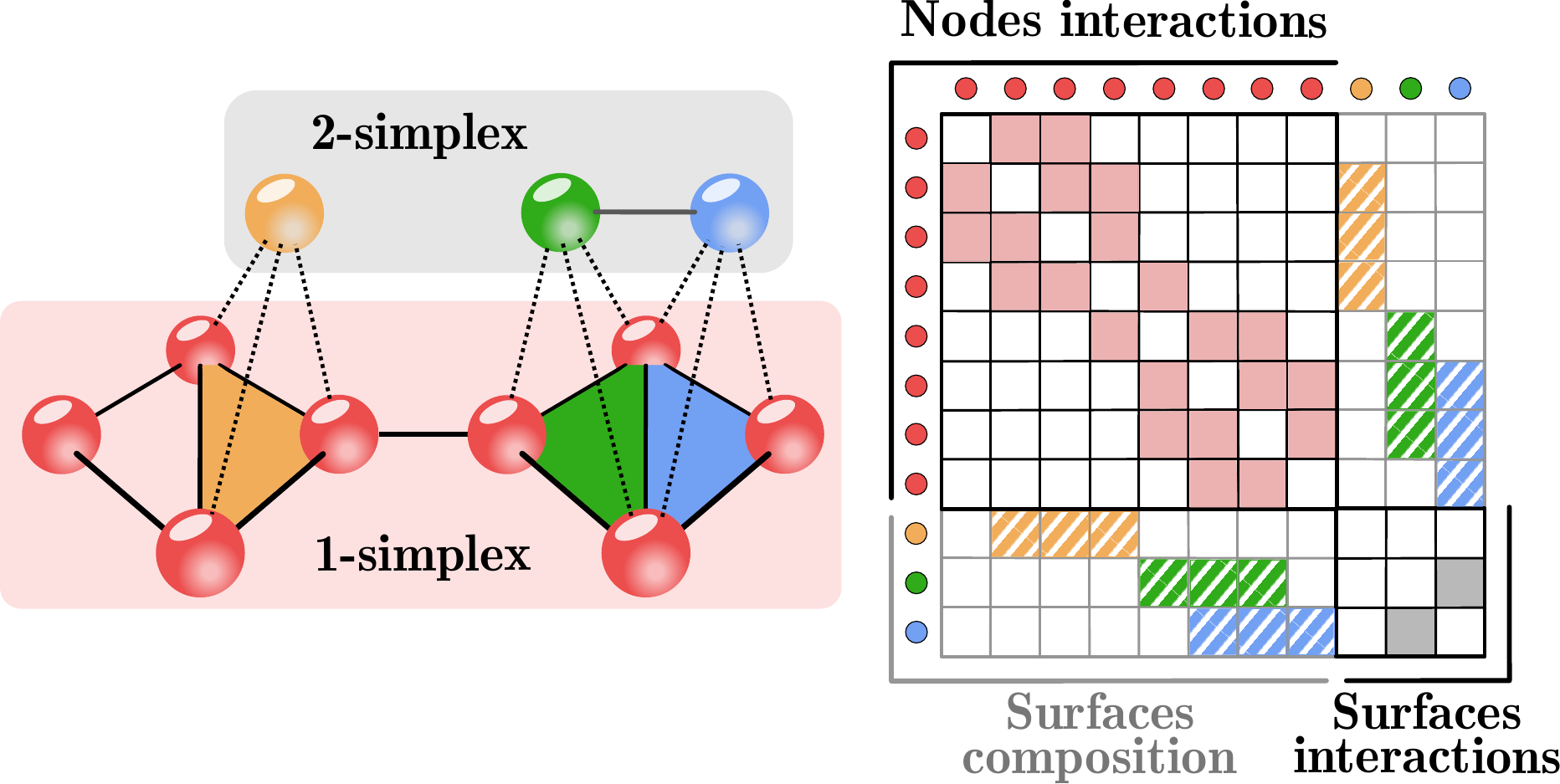}
    \caption{\textbf{From a simplicial complex to a two-layer network.} 
Left:    
    a simplicial complex of eight nodes with three filled triangles (2-simplices, colored), and its two-layer representation used by HHO. The first layer (bottom, red nodes) is the 1-skeleton, whose links are the pairwise interactions. The second layer (top) contains one node per filled triangle, linked to its three vertices by inter-layer links (dotted lines); the green and blue triangles share an edge and are therefore linked within the second layer. 
Right: the corresponding supra-adjacency matrix $\mathcal{A}$, equation~\eqref{eq:supra}. The node--node block (nodes interactions) is the weighted adjacency matrix $\widetilde{W}$ of the 1-skeleton; the off-diagonal blocks (surfaces composition, hatched) are the triangle--node incidence matrix weighted by $\mu$; the triangle--triangle block (surfaces interaction, gray) is $S$, with entries $2\gamma$ for triangles sharing an edge. The two-layer network is embedded in the Poincar\'e disk; radii depend only on the strength in the first layer, so the higher-order information acts on the angular coordinates only.}
    \label{mat_cons}
\end{figure}

Figure~\ref{mat_cons} 
illustrates the main 
idea of the HHO framework.  
An embedding based solely on pairwise interactions cannot distinguish a group of three nodes participating in a higher-order interaction from three nodes connected exclusively through pairwise interactions, as both configurations
produce 
the same triangle in the projected network. HHO 
resolves 
this ambiguity through a multilayer network representation~\cite{Boccaletti2014}. The first layer consists of the original nodes, connected by links representing pairwise interactions. 
The second layer represents higher-order interactions, with each filled triangle represented by a distinct node and two such nodes connected whenever the corresponding triangles share an edge.
Inter-layer links connect each triangle node to its three constituent vertices, forming a bipartite incidence graph. Unlike multiplex networks, in which different layers typically represent the same set of nodes, HHO comprises two layers with distinct node sets, explicitly distinguishing individual entities from higher-order interactions. When the resulting two-layer network is embedded in the hyperbolic disk using our multilayer embedding method~\cite{GuillemaudMultilayer2025}, each triangle node draws its three constituent vertices towards a common angular direction, while preserving their radial coordinates, which reflect their local connectivity strength. The triangle nodes therefore act as a geometric scaffold that shapes the angular organization of the original nodes, whose embedded positions are subsequently analyzed. The hyperbolic area of the triangle defined by the three members of a group thus provides a quantitative measure of their geometric concentration, with smaller areas indicating a more compact representation of the corresponding higher-order interaction.

\subsection{Simplicial complexes}\label{Sec:method_sc}
A simple undirected graph $G=(V,E)$ consists of a set $V=\{v_1,v_2,\dots,v_N\}$ of $N$ nodes (or vertices) and a set $E$ of edges, each of which is an unordered pair of distinct nodes. Assigning a positive weight to each edge gives the weighted adjacency matrix $W\in\mathbb{R}^{N\times N}$, where $w_{ij}=w_{ji}>0$ if $\{v_i,v_j\}\in E$ and $w_{ij}=0$ otherwise. For an unweighted graph, every edge has weight $1$. The degree $k_i$ of node $v_i$ is its number of neighbors, and its strength is $s_i=\sum_{j=1}^{N}w_{ij}$~\cite{Latora2017}. Edges encode pairwise interactions only and cannot represent collective interactions among three or more nodes.

A simplicial complex $\mathcal{K}$ with node set $V$ is a collection of non-empty subsets of $V$, called simplices, 
and is closed under inclusion: if $\sigma\in\mathcal{K}$ and $\tau\subseteq\sigma$ is non-empty, then $\tau\in\mathcal{K}$~\cite{Battiston2020,Bianconi2021}. A simplex $\sigma=\{v_0,\dots,v_k\}$ containing $k+1$ vertices has dimension $k$ and is called a {$k$-simplex}.
In a simplicial complex $\mathcal{K}$, each non-empty subset of a simplex is a {face}, including the simplex itself; a {proper face} is a face strictly contained in that simplex. Thus, $0$-simplices are nodes, $1$-simplices are edges, $2$-simplices are filled triangles, and $3$-simplices are filled tetrahedra. For $k\geq2$, a $k$-simplex has $\binom{k+1}{3}$ two-dimensional faces, each corresponding to a choice of three vertices. A simplex is maximal if it is not a proper face of any other simplex in $\mathcal{K}$.
For example, an article co-authored by three authors $a,b,c$ can be represented by the $2$-simplex $\{a,b,c\}$. Closure under taking non-empty subsets requires that the three edges $\{a,b\}$, $\{a,c\}$, and $\{b,c\}$, together with the three vertices $\{a\}$, $\{b\}$, and $\{c\}$, also belong to $\mathcal{K}$. The converse does not hold: the presence of all three edges does not imply that the filled triangle belongs to $\mathcal{K}$. This closure requirement distinguishes a simplicial complex from a general hypergraph, in which subsets of a hyperedge need not themselves be hyperedges~\cite{Battiston2020,Bianconi2021}.

The 1-skeleton of $\mathcal{K}$ is the graph $G=(V,E)$ formed by its $0$- and $1$-simplices, with adjacency matrix $W$ (in the co-authorship complex, for instance, $w_{ij}$ is the number of articles co-signed by $i$ and $j$). Every $2$-simplex of $\mathcal{K}$ appears in $G$ as a triangle, i.e.\ a 3-clique. We call a triangle of $G$ \emph{filled} if it is a face of a simplex of dimension $k\geq2$, i.e.\ if its three nodes take part in a common group interaction, and \emph{empty} otherwise, i.e.\ if it results from three separate pairwise interactions. Two complexes can therefore have the same 1-skeleton and differ only in which of its triangles are filled. Unlike the clique complex of a graph, which fills every clique by construction\cite{Giusti2016}, the complexes analyzed here take their filled simplices from the data, so they carry information that is not contained in $G$ and is invisible to any method that only reads $G$. In what follows, ``complex'' without qualifier denotes a simplicial complex; complexes of proteins are always called protein complexes. We write $\mathcal{S}$ for the set of higher-order simplices (of dimension $k\geq 2$) given in the data, and $\mathcal{T}=\{\tau_1,\dots,\tau_M\}$ for the set of filled triangles, i.e.\ of the distinct triangular faces of the simplices of $\mathcal{S}$.

\subsection{From a simplicial complex to a two-layer network}\label{Sec:method_ml}
HHO recasts $\mathcal{K}$ as a two-layer network (figure~\ref{mat_cons}). The first layer is the 1-skeleton $G$ of $\mathcal{K}$. The second layer contains one node per filled triangle $\tau_a\in\mathcal{T}$: a $k$-simplex of $\mathcal{S}$ contributes its $\binom{k+1}{3}$ triangular faces, and a triangle that is a face of several simplices is a single node. Let $B\in\{0,1\}^{M\times N}$ be the triangle--node incidence matrix, $B_{ai}=1$ if $i\in\tau_a$, so that $(BB^{\top})_{ab}=|\tau_a\cap\tau_b|$ counts the nodes shared by two triangles. The supra-adjacency matrix of the two-layer network is
\begin{equation}\label{eq:supra}
\mathcal{A}=\begin{pmatrix} \widetilde{W} & \mu\,B^{\top}\\ \mu\,B & S\end{pmatrix},\qquad
S_{ab}=\begin{cases}\gamma\,|\tau_a\cap\tau_b| & \text{if } a\neq b \text{ and } |\tau_a\cap\tau_b|>1,\\ 0 & \text{otherwise.}\end{cases}
\end{equation}
The inter-layer coupling $\mu>0$ links each triangle node to its three vertices, and the intra-layer coupling $\gamma\geq 0$ links two triangles that share an edge ($|\tau_a\cap\tau_b|=2$). The node--node block $\widetilde{W}$ is either $W$ itself or its repulsion--attraction (RA) pre-weighted version, as proposed in Ref.~\cite{Muscoloni2017},
\begin{equation}\label{eq:ra}
w^{\mathrm{RA}}_{ij}=\frac{k_i+k_j+k_i k_j}{1+c_{ij}},\qquad \widetilde{W}_{ij}=\log\!\left(1+w^{\mathrm{RA}}_{ij}\right)\quad\text{for } (i,j)\in E,
\end{equation}
where $k_i$ is the (unweighted) degree of node $i$ and $c_{ij}$ the number of common neighbors of $i$ and $j$.

We use the filled triangles, rather than the simplices themselves, as the nodes of the second layer for two reasons. First, every triangle node is coupled to its three vertices with the same strength $\mu$, whether the triangle belongs to one simplex or to several and whatever their dimension; a $k$-simplex thus acts through its $\binom{k+1}{3}$ triangle nodes. Second, the triangles are the objects whose hyperbolic areas we use to evaluate the embeddings (section~\ref{Sec:method_area}), so that the second layer acts directly on the quantities that are measured. On the synthetic complexes, whose simplices are all triangles, $\mathcal{T}=\mathcal{S}$. Simplices of higher dimension thus enter the embedding through their triangular faces (see section~\ref{Sec:discussion} for an extension with one layer per dimension).

\subsection{Hyperbolic embedding}\label{Sec:method_emb}
HHO models the simplicial complex as the two-layer network of section~\ref{Sec:method_ml}, and embeds this network in the Poincar\'e disk $\mathbb{D}=\{x\in\mathbb{R}^2:\|x\|<1\}$, of curvature $-1$, with our multilayer hyperbolic embedding\cite{GuillemaudMultilayer2025}, which follows the coalescent-embedding principle of separating the angular and the radial coordinates\cite{Muscoloni2017}. The angular coordinates are obtained from the supra-adjacency matrix, mapped element-wise to the dissimilarity $D=\exp(-\mathcal{A}/\beta_{\mathrm{dist}})$, so that unlinked pairs ($\mathcal{A}_{ij}=0$) receive the maximal dissimilarity $D_{ij}=1$ and linked pairs have a dissimilarity that decreases with the strength of their link; the diagonal of $D$ plays no role. Matrix $D$ contains thus the connectivity information used by the dimensionality reduction method Isomap~\cite{Tenenbaum2000}, as in the original coalescent embeddings~\cite{Muscoloni2017} and in our multilayer embedding approach~\cite{GuillemaudMultilayer2025}. Each embedded node $i$ receives the angle $\theta_i$ of a two-dimensional coordinate. Because the triangle nodes take part in the dimensionality reduction method, each filled triangle attracts its three vertices towards a common angular sector; nodes of the second layer are used only to shape the angular organization of the first ones.

The radial coordinate, by contrast, depends only on the strength $s_i=\sum_j w_{ij}$ of node $i$ in the 1-skeleton,
\begin{equation}\label{eq:radius}
r_i = 1-\tanh\!\left(s_i/\beta\right),
\end{equation}
where $w_{ij}$ are the weights of $G$ (the weights $w^{\mathrm{RA}}_{ij}$ of equation~\eqref{eq:ra} when pre-weighting is used) and $\beta$ sets the radial scale. High-strength nodes are placed close to the center of the disk, so that their radial coordinate encodes their local connectivity~\cite{Krioukov2010,Papadopoulos2012}. Embedding coordinates are thus $x_i=r_i(\cos\theta_i,\sin\theta_i)$ and distances are geodesic in $\mathbb{D}$ defined by,
\begin{equation}\label{eq:dist}
d_{\mathbb{H}}(x_i,x_j)=\operatorname{arccosh}\!\left(1+\frac{2\|x_i-x_j\|^2}{(1-\|x_i\|^2)(1-\|x_j\|^2)}\right).
\end{equation}

\subsection{Hyperbolic area of triangles as a marker of higher-order organization}\label{Sec:method_area}
Three embedded nodes $x_i,x_j,x_k$ define a geodesic triangle in $\mathbb{D}$, whose sides are arcs of the geodesics joining them, with lengths $a=d_{\mathbb{H}}(x_j,x_k)$, $b=d_{\mathbb{H}}(x_i,x_k)$ and $c=d_{\mathbb{H}}(x_i,x_j)$ given by equation~\eqref{eq:dist}. By the Gauss--Bonnet theorem, a geodesic triangle in a space of constant curvature $-1$ has an area equal to its angular defect\cite{Beardon1983,Ratcliffe2019},
\begin{equation}\label{eq:area}
A_{ijk}=\pi-(\alpha_i+\alpha_j+\alpha_k),
\end{equation}
where $\alpha_i,\alpha_j,\alpha_k$ are its interior angles. The angles follow from the side lengths through the hyperbolic law of cosines\cite{Beardon1983,Ratcliffe2019},
\begin{equation}\label{eq:cosines}
\cos\alpha_i=\frac{\cosh b\,\cosh c-\cosh a}{\sinh b\,\sinh c},
\end{equation}
and similarly for $\alpha_j$ and $\alpha_k$ by permutation of the sides. Unlike a Euclidean triangle, whose area grows without bound with its sides, a hyperbolic triangle has an area bounded by $\pi$, which is approached when the three vertices move towards the boundary of the disk far from each other. The area vanishes when the three nodes are aligned on a geodesic or collapse onto a single point. In practice, the area is a single scalar that measures how tightly the three members of a triangle are gathered in the latent space, taking both their angular separation and their radial positions into account. 

Here, we use three markers, computed from the coordinates of the embedded nodes only.
\begin{enumerate}
\item \emph{Area ratio} (HHO versus pairwise embedding). For each filled triangle $t$, we compute $A_t^{\mathrm{HHO}}/A_t^{\mathrm{pairwise}}$, the ratio of its area under HHO to its area under the pairwise embedding. A ratio below $1$ means that the triangle has been made more compact by the higher-order information. We report the median ratio and the fraction of triangles with a ratio below $1$, and test whether the paired areas differ with a Wilcoxon signed-rank test\cite{Wilcoxon1945}.
\item \emph{Filled versus empty triangles}. To assess whether the HHO embedding captures higher-order organization, we quantify the separation between filled and empty triangles in terms of their hyperbolic areas. Although both types of triangles are formed by nodes of the same underlying graph $G$, an embedding that effectively captures group interactions is expected to yield smaller areas for filled triangles, which represent higher-order interactions, than for empty triangles, which do not. We characterize this separation using three complementary measures. First, we compute the area fold, defined as the ratio of the median area of empty triangles to that of filled triangles, $\mathrm{median}(A_{\mathrm{empty}})/\mathrm{median}(A_{\mathrm{filled}})$, with values greater than 1 indicating smaller filled triangles. Second, we quantify the probability of pairwise area separation, $P(A_{\mathrm{filled}}<A_{\mathrm{empty}})$, defined as the probability that a randomly selected filled triangle has a smaller area than a randomly selected empty triangle. This probability is computed exactly over all filled--empty triangle pairs, with ties contributing 1/2, and is equivalent to the area under the receiver operating characteristic curve (AUC) when triangle area is used to discriminate filled from empty triangles. Accordingly, an AUC of 0.5 indicates no discriminatory ability, whereas an AUC of 1 corresponds to perfect separation. Third, we assess the statistical significance of the difference in area distributions using the two-sided Mann--Whitney $U$ test~\cite{MannWhitney1947}, whose test statistic is directly related to the pairwise ranking probability. Together, these measures provide complementary assessments of the magnitude, discriminatory performance, and statistical significance of the separation: the area fold reflects differences in the scale of the distributions, whereas the AUC depends only on the relative ranking of triangle areas.
\item \emph{Global distance preservation}. Compacting the higher-order groups should not come at the cost of the overall organization of the network. We therefore measure how well embedding distances preserve graph distances by the Spearman rank correlation $\rho$ between the hop distance (the shortest path length) in the 1-skeleton $G$ and the hyperbolic distance $d_{\mathbb{H}}$, over all pairs of nodes. A value of $\rho$ close to $1$ means that nodes that are close in the original network are also close in the disk; a value close to $0$ means that the embedding does not preserve the network distances.
\end{enumerate}
Taken together, a small area ratio indicates that the embedding gathers the members of the higher-order interactions in a compact region of the latent space, a separation between filled and empty triangles indicates that this compactness is specific to genuine group interactions in the simplicial network rather than to its pairwise 1-skeleton, and the distance preservation $\rho$ indicates whether the global geometry is preserved. On the synthetic, Cora and \textit{E.~coli} simplicial complexes, the filled triangles are also those used to build the second layer of HHO, so the first two markers measure how faithfully HHO represents the information it is given rather than an independent prediction.

\subsection{Evaluation criteria and statistics}\label{Sec:method_stats}
Besides the area markers, we evaluate group recovery (C1) for each node, defined as the fraction of its $m$ nearest embedded neighbors that are partners, where partners are co-members of a higher-order simplex (e.g. co-authors), or of an annotated protein complex in the \textit{E.~coli} network. Here, we set $m=5$ to provide a compromise between local specificity and neighborhood representativeness, ensuring that group recovery reflects the concentration of partners in the immediate embedding neighborhood while remaining sufficiently robust to the influence of individual nearest neighbors. We further report the edge versus non-edge AUC given by $P(d_{ij}^{\mathrm{edge}}<d_{kl}^{\mathrm{non\text{-}edge}})$.

We compare HHO with four methods, all producing two-dimensional embeddings. They differ in two respects: whether they use the higher-order simplices or only the 1-skeleton $G$, and whether they embed in the Poincar\'e disk or in the Euclidean plane.
\begin{enumerate}
\item \emph{Pairwise embedding} (hyperbolic, pairwise). Our multilayer hyperbolic embedding~\cite{GuillemaudMultilayer2025} restricted to a single layer, i.e.\ a coalescent-type embedding of the 1-skeleton~\cite{Muscoloni2017}: the procedure of section~\ref{Sec:method_emb} applied to $G$ alone, i.e.\ with $\mathcal{A}=\widetilde{W}$ and no second layer. It is the reference of the area ratio marker described above. 
\item \emph{Weighted clique expansion} (hyperbolic, higher-order). This is the hyperbolic embedding based on weighted clique expansion~\cite{Zien1999,Agarwal2006,Newman2001}, in which the edge weights of the 1-skeleton are augmented to account for higher-order interactions. Specifically, the weight of each edge $(i,j)$ in the matrix used to determine the angular coordinates is increased by the number of simplices in $\mathcal{S}$ that contain both $i$ and $j$. All other steps follow the pairwise embedding procedure, including the assignment of radial coordinates based on the same node strengths as in HHO. This comparison allows us to assess whether explicitly representing higher-order interactions through the two-layer construction provides an advantage over incorporating this information through edge reweighting alone.
\item \emph{Laplacian eigenmaps}~\cite{BelkinNiyogi2003} (Euclidean, pairwise). As a Euclidean baseline, we consider Laplacian eigenmaps applied to the weighted pairwise graph defined by the adjacency matrix $W$. The node coordinates are obtained from the eigenvectors associated with the two smallest non-zero eigenvalues of the normalized graph Laplacian, yielding a two-dimensional Euclidean embedding that captures the large-scale connectivity structure of the graph. This approach provides a reference for assessing the extent to which the proposed HHO embedding captures higher-order organization beyond that represented by a pairwise graph embedded in Euclidean space.
\item \emph{Hypergraph spectral embedding}~\cite{Zhou2006} (Euclidean, higher-order). The simplices of $\mathcal{S}$ and the edges of $G$ are treated as hyperedges of unit weight, and the coordinates are the eigenvectors of the second and third smallest eigenvalues of the normalized hypergraph Laplacian $\Delta=I-D_v^{-1/2}HD_e^{-1}H^{\top}D_v^{-1/2}$, where $H$ denotes the node--hyperedge incidence matrix, and $D_v$, $D_e$ the node and hyperedge degree matrices (eigenvectors unscaled), respectively.
\end{enumerate}

The benchmark set therefore comprises two Euclidean embedding methods, one based on the pairwise 1-skeleton $G$ and another based on higher-order interactions through the simplices. For these Euclidean embeddings, triangle areas are computed in the plane. Because planar and hyperbolic areas have different geometric properties and scales, their absolute values are not directly comparable. Accordingly, cross-geometry comparisons of triangle areas are restricted to the area-based AUC, which quantifies the relative ranking of filled and empty triangles and is invariant under strictly increasing transformations of the area measure. This enables a consistent assessment of the ability of Euclidean and hyperbolic embeddings to distinguish higher-order interactions from triangles formed solely by pairwise connections.

Two kinds of statistical tests are used. Within the analysis of each of the studied complexes (sections~\ref{Sec:res_synth}--\ref{Sec:res_ecoli}), the paired areas of the pairwise and HHO embeddings are compared with the Wilcoxon signed-rank test over filled triangles, and filled and empty triangles with the Mann--Whitney $U$ test (section~\ref{Sec:method_area}). In the comparison of the embedding methods (section~\ref{Sec:res_comparison}), the unit of replication depends on the dataset. For the Holme--Kim model, we generate 10 independent graphs, average each criterion over the 10 graphs and compare the methods with paired Wilcoxon tests across graphs. For each real network, 95\% confidence intervals are obtained by bootstrap over the marker estimated for each node~\cite{Efron1979}: nodes are redrawn at random with replacement, as many times as there are nodes, and the marker is recomputed (averaged) on each resample. Group recovery C1 is compared with paired Wilcoxon tests over nodes, and the area AUCs with a paired bootstrap (2000 resamples), in which filled and empty triangles are redrawn in the same way and the same resamples are used for every method. 

The Erd\H{o}s--R\'enyi complex, which serves as a counter-example without hyperbolic structure (section~\ref{Sec:data_synth}), is analyzed with the area markers and $\rho$ only and is not part of the comparison. We report uncorrected two-sided $p$-values. Each comparison of HHO with another method answers a separate question, and a claim that HHO is above all the other methods holds only if every one of the corresponding tests rejects; such an intersection--union test has level $\alpha$ without correction for multiplicity~\cite{Berger1982}. The smallest attainable $p$-value is $0.002$ with 10 graphs (exact Wilcoxon test) and $0.001$ with 2000 resamples, and we write $p<10^{-300}$ when a $p$-value is below the double-precision range. Tests over triangles, which share vertices, and over nodes of a single network treat dependent nodes as independent observations and therefore overstate the evidence; we report them for completeness, but we base our conclusions on effect sizes and confidence intervals. 

\section{Data}\label{Sec:data}

We study four simplicial complexes: two synthetic complexes based on Holme--Kim and Erd\H{o}s--R\'enyi graphs, and two empirical complexes representing scientific collaborations in Cora and protein interactions in \textit{E.~coli}. Table~\ref{tab:datasets} summarizes the structural properties of their giant connected components, including size, degree statistics, clustering, and the numbers of filled and empty triangles. The synthetic complexes have similar numbers of triangles, approximately half of which are filled, but differ markedly in clustering and maximum degree. The empirical complexes contain many more triangles, most of which are empty. 

\begin{table}[!htbp]
\caption{
Structural properties of the 
simplicial complexes studied here: number of nodes $N$, number of edges $|E|$ in the 1-skeleton, mean and maximum node degree 
$\langle k\rangle$ and $k_{\max}$, 
average clustering coefficient $C$,  
3-cliques of the 1-skeleton (triangles), classified as filled if they belong to the simplicial complex as 2-simplices, and empty otherwise. Filled triangles define the nodes of the second HHO layer. For Cora, they comprise the triangular faces of 179, 61, 28, and 13 simplices of dimensions 2, 3, 4, and 5, respectively. For \textit{E.~coli}, they comprise the triangular faces of the maximal cliques associated with 14 Complex Portal complexes. Groups lists the node annotations of the empirical complexes: the scientific fields of Cora and the COG functional categories of \textit{E.~coli} color the nodes in figures~\ref{fig:cora} and~\ref{fig:ecoli}, and the protein complexes define the simplices of \textit{E.~coli}.
}
\label{tab:datasets}
\centering
\small
\adjustbox{max width=\textwidth}{%
\begin{tabular}{lrrrrrrrrl}
\hline
dataset & $N$ & $|E|$ & $\langle k\rangle$ & $k_{\max}$ & $C$ & triangles & filled & empty & groups \\
\hline
Holme--Kim & 500 & 996 & 3.98 & 61 & 0.228 & 178 & 87 & 91 & -- \\
Erd\H{o}s--R\'enyi & 500 & 2571 & 10.28 & 20 & 0.019 & 175 & 87 & 88 & -- \\
Cora & 723 & 2243 & 6.20 & 60 & 0.542 & 5259 & 834 & 4425 & 7 fields \\
\textit{E.~coli} & 286 & 5027 & 35.15 & 103 & 0.696 & 74{,}918 & 5366 & 69{,}552 & 14 protein complexes, 18 COG categories \\
\hline
\end{tabular}}
\end{table}

\subsection{Synthetic complexes with planted surfaces}\label{Sec:data_synth}
We generated graphs from two models: Holme--Kim\cite{HolmeKim2002} (powerlaw-cluster graph, $N=500$, two edges per new node, triangle-formation probability $0.3$) and Erd\H{o}s--R\'enyi\cite{ErdosRenyi1959,Gilbert1959} ($N=500$, connection probability $0.02$, so that it has as many nodes as the Holme--Kim graph and about as many triangles). To obtain a simplicial complex, 
each triangle of the graph was independently declared filled (a 2-simplex) with probability $50/101\approx0.495$ (a uniform integer draw in $\{0,\dots,100\}$ below 50), and left empty otherwise.
We analyze the giant component of each graph. Holme--Kim is our main model because it combines a heavy-tailed degree distribution with high clustering, the two properties that a latent hyperbolic geometry reproduces\cite{Krioukov2010}, and hence has many triangles of both kinds. The Erd\H{o}s--R\'enyi graph, with a Poisson degree distribution and a clustering set by its density alone, has neither property and serves as a counter-example without hyperbolic structure; the single-graph results use one graph per model (graph seeds 1000 for Holme--Kim and 103 for Erd\H{o}s--R\'enyi, embedding seed 0), and the comparison of section~\ref{Sec:res_comparison} uses 10 further Holme--Kim graphs (seeds 0--9).

\subsection{Co-authorship complex (Cora)}\label{Sec:data_cora}
The Cora co-authorship hypergraph, derived from the Cora collection of computer-science articles\cite{McCallum2000}, is a standard benchmark for hypergraph learning, in which articles are vertices and authors hyperedges\cite{Yadati2019}. We use the dual representation distributed by the AHORN repository (revision 2, Zenodo record 19232933), in which nodes are authors and each article is a simplex on its authors; each author carries one of 7 scientific-field labels, provided by AHORN as a node attribute (the rule by which a field is assigned to an author is not documented in the repository). The 1-skeleton is built from all articles, with edge weights equal to the number of articles co-signed by a pair, and articles with 3 to 6 authors (dimension $2\leq k\leq 5$) form the higher-order simplices. The giant component has 723 authors and 2243 edges, with 179, 61, 28 and 13 simplices of dimension 2, 3, 4 and 5, respectively. These simplices contain 834 distinct filled triangles, which form the second layer of HHO.

\subsection{Protein complexes of \textit{Escherichia coli}}\label{Sec:data_ecoli}
Protein--protein associations of \textit{E.~coli} K-12 MG1655 were taken from STRING v12.0\cite{Szklarczyk2023} and thresholded at a combined score of 700 (high confidence), which gives 4006 proteins and 32{,}244 interactions. We restricted the analysis to the 4-core of this graph, the maximal subgraph in which every protein has at least four partners (3038 proteins and 30{,}321 interactions). Because the two-layer embedding uses a dense $(N+M)\times(N+M)$ matrix, we analyze a subnetwork of 300 proteins of the 4-core, obtained by a deterministic rule that does not use the complex annotations: proteins are taken in the order in which they first appear in the STRING association file, which is sorted by locus tag, i.e.\ by position on the chromosome. The subnetwork is therefore formed by the 4-core proteins in the high-confidence neighborhoods of the first genes of the chromosome map, from the \textit{thr} operon (\textit{thrL}, b0001) through the chaperones \textit{dnaK} and \textit{dnaJ} to the ribosomal protein gene \textit{rpsT} (b0023), within whose neighborhood the bound of 300 proteins is reached. Since this neighborhood is dominated by ribosomal, translation and transcription proteins (63 of the 99 selected proteins it contains), the subnetwork contains a dense block of the transcription and translation machinery, and we treat it as a case study of large multi-protein assemblies rather than as a sample of the \textit{E.~coli} complexome. It has three connected components, of 286, 13 and 1 proteins; we analyze the giant component, with 286 proteins and 5027 interactions.

Protein complexes were taken from the manually curated Complex Portal\cite{Meldal2022}, which includes a draft of the \textit{E.~coli} complexome, and mapped to STRING identifiers through UniProt accessions. Complexes sharing a name were merged, and 160 complexes have at least three members mapped to the \textit{E.~coli} STRING network; restricted to the analyzed giant component, 14 complexes keep at least three members, and all of them contribute simplices.
For each protein complex, the higher-order simplices are the maximal cliques of size $\geq 3$ of the subgraph of the analyzed network induced by its members, and the filled triangles, which form the second layer of HHO, are the triangles contained in one of these cliques. The 14 protein complexes cover 64 proteins: the two ribosomal subunits, the sigma-factor variants of the RNA polymerase holoenzyme (which share their core subunits), two overlapping Sec translocon complexes, the RNA degradosome, and two transcription elongation and antitermination complexes.
The giant component contains 5366 filled and 69{,}552 empty triangles; 206 empty triangles, in which two vertices coincide to numerical precision in both embeddings, have an undefined area and are excluded from the area markers. Seventy proteins share a complex with at least one other protein of the giant component (the 64 proteins in simplices and 6 whose complex forms no triangle in it). Here, each protein was assigned the first functional category of its Cluster of Orthologous Genes (COG) annotation~\cite{Galperin2021} (18 categories, 11 proteins of unknown category; the largest category is translation, ribosomal structure and biogenesis, with 84 proteins).

\begin{figure}[!htbp]
\centering
\includegraphics[width=0.9\textwidth]{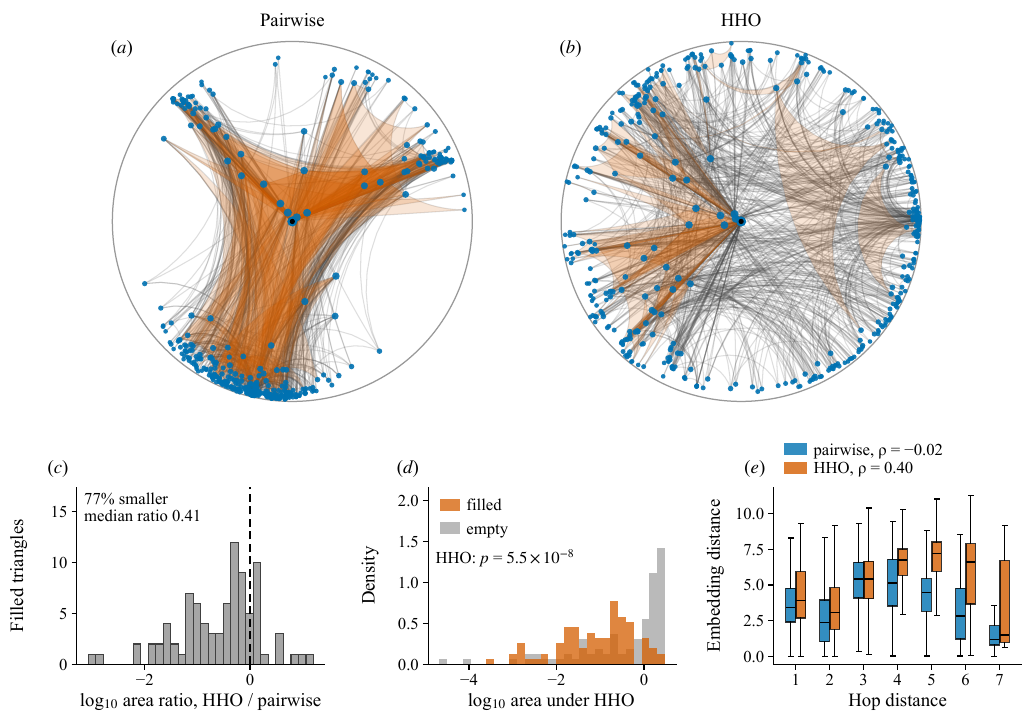}
\caption{
HHO reduces the hyperbolic area of filled triangles in a synthetic Holme--Kim complex and distinguishes them geometrically from empty triangles. 
Results for a single   Holme--Kim graph with planted filled triangles (graph seed 1000, embedding seed 0). 
Its giant connected component contains 500 nodes,  
996 edges, 87 filled triangles and 91 empty triangles. (\textit{a}),  (\textit{b}), 
Pairwise embedding of the 1-skeleton and HHO embedding in the Poincar\'e disk, respectively.  Shaded regions indicate  filled triangles, gray curves show edges (geodesics), and the black dot marks the center of the disk. 
Node marker area is proportional to $\sqrt{\mathrm{degree}+1}$.
(\textit{c}) Distribution of $\log_{10}(A_{\mathrm{HHO}}/A_{\mathrm{pairwise}})$ for the 87 filled triangles. The dashed line at zero indicates equal areas.
The median area ratio is 0.406, and 77\% of the filled triangles have 
smaller area under HHO (paired Wilcoxon $p=4.6\times10^{-7}$). (\textit{d}) Densities of $\log_{10}$ hyperbolic areas of filled (87) and empty (91) triangles under HHO; fold $\mathrm{median}(A_{\mathrm{empty}})/\mathrm{median}(A_{\mathrm{filled}})=8.28$, $P(A_{\mathrm{filled}}<A_{\mathrm{empty}})=0.736$, Mann--Whitney $p=5.5\times10^{-8}$. Pairwise embedding (not drawn): fold 1.57, $P=0.543$, $p=0.32$. (\textit{e}) Embedding distance against shortest path distance over all node pairs (boxes: quartiles and median; whiskers: 1.5 times the interquartile range; outliers not drawn). Spearman $\rho$: pairwise $-0.025$, HHO 0.402. All numbers come from a single graph; the mean over 10 graphs is in table~\ref{tab:comparison}.}
\label{fig:synth}
\end{figure}

\section{Results}\label{Sec:results}
\subsection{Synthetic complexes: HHO compacts planted surfaces}\label{Sec:res_synth}
We first embedded a Holme--Kim graph with planted surfaces, whose giant component has 500 nodes and 996 edges, with 87 filled and 91 empty triangles (figure~\ref{fig:synth}).
Figure~\ref{fig:synth}\textit{a,b} shows the pairwise and HHO embeddings of this complex, with the filled triangles drawn as shaded regions.
Under HHO, 77\% of the filled triangles are smaller than under the pairwise embedding, with a median area ratio of 0.406 (paired Wilcoxon $p=4.6\times10^{-7}$; figure~\ref{fig:synth}\textit{c}).
Within the HHO embedding, filled triangles are smaller than empty ones (fold 8.28, $P(A_{\mathrm{filled}}<A_{\mathrm{empty}})=0.736$, Mann--Whitney $p=5.5\times10^{-8}$; figure~\ref{fig:synth}\textit{d}), whereas the pairwise embedding does not separate them (fold 1.57, $P=0.543$, $p=0.32$).
On this graph, HHO distances follow shortest path lengths ($\rho=0.402$), whereas those of the pairwise embedding do not ($\rho=-0.025$; figure~\ref{fig:synth}\textit{e}).

As a counter-example to networks with latent hyperbolic structure, we applied HHO to Erd\H{o}s--R\'enyi graphs. HHO yields smaller areas for most filled triangles than the corresponding triangles in the pairwise embedding (98.9\% have an area ratio below 1; median ratio 0.0137; paired Wilcoxon $p=6.1\times10^{-16}$) and separates them clearly from the empty triangles. Specifically, HHO achieves an area fold of 65.3 and a pairwise ranking probability $P(A_{\mathrm{filled}}<A_{\mathrm{empty}})=0.952$ (Mann--Whitney $p=5.5\times10^{-25}$). For the pairwise embedding, the corresponding values are $1.08$, $0.550$, and $p=0.254$, respectively. However, HHO does not improve global distance preservation ($\rho=0.244$, compared with $0.302$ for the pairwise embedding), on this graph and on 10 further ones (seeds 0--9). These results suggest that, although incorporating higher-order information can locally compact filled triangles and separate them from empty ones, this effect does not translate into an improvement in the global representation. This behavior is consistent with the absence of an underlying latent geometry in Erd\H{o}s--R\'enyi graphs, where edges are generated independently with a common probability, producing a relatively homogeneous degree distribution and clustering attributable to chance. Consequently, such networks lack the pronounced radial hierarchy and angular similarity structure typically associated with hyperbolic organization, and the higher-order interactions provide no systematic geometric scaffold on which to organize the embedded nodes.

\begin{figure}[!htbp]
\centering
\includegraphics[width=0.9\textwidth]{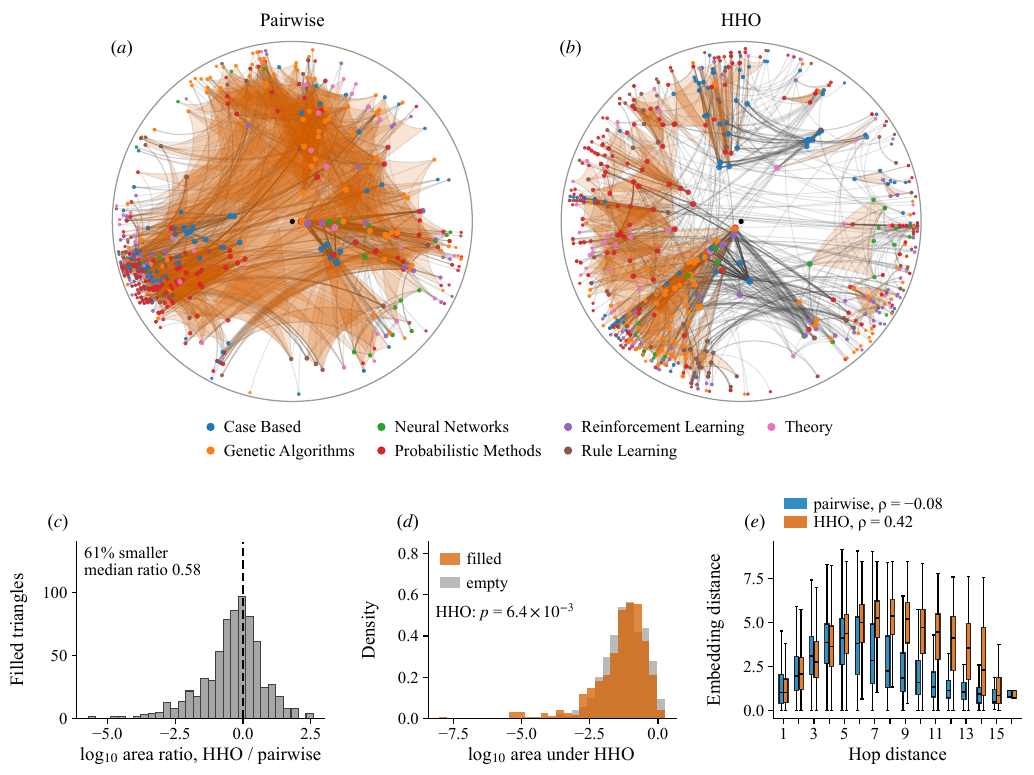}
\caption{\textbf{
HHO reduces the area of most filled triangles and improves the correspondence between hyperbolic and graph distances in the Cora co-authorship complex, but provides only weak separation between filled and empty triangles.
}
The giant connected component contains 723 authors and 2243 edges, with 179, 61, 28, and 13 higher-order simplices of dimensions 2, 3, 4, and 5, respectively.
(\textit{a}) Pairwise and (\textit{b}) HHO embeddings. 
Node colors indicate the seven scientific fields provided by AHORN. 
Shaded regions are filled triangles, gray curves the edges and the black dot marks the center of the disk, as in figure~\ref{fig:synth}. (\textit{c})
Distribution of $\log_{10}(A_{\mathrm{HHO}}/A_{\mathrm{pairwise}})$ for the 706 filled triangles 
with a defined area (three distinct vertices) in both embeddings; the 704 with a non-zero HHO area are drawn, the other two having zero area to numerical precision.
The dashed line at zero indicates equal areas. Median ratio 0.577, 61.3\% smaller under HHO; paired Wilcoxon $p=9.2\times10^{-18}$. (\textit{d}) Densities of $\log_{10}$ areas of filled (704 of 706; see (\textit{c})) and empty (3093) triangles under HHO: fold 1.05, $P(A_{\mathrm{filled}}<A_{\mathrm{empty}})=0.533$, Mann--Whitney $p=0.0064$. Pairwise embedding (not drawn; 709 filled, 3093 empty): fold 0.236, $P=0.316$, $p=3.9\times10^{-53}$, i.e.\ under the pairwise embedding the ordering is inverted. (\textit{e}) Embedding distance against shortest path distance (boxes as in figure~\ref{fig:synth}\textit{e}; every second hop is labeled). Spearman $\rho$: pairwise $-0.0784$, HHO 0.417.}
\label{fig:cora}
\end{figure}

\subsection{Co-authorship complex: higher-order structure reorganizes the angular coordinates}\label{Sec:res_cora}
On the Cora complex (723 authors, 2243 edges; 179, 61, 28 and 13 simplices of dimension 2 to 5), figure~\ref{fig:cora}\textit{a,b} shows the pairwise and HHO embeddings. Visually, HHO separates the groups more clearly. The filled triangles, which span most of the disk under the pairwise embedding, are confined to localized angular sectors under HHO, and the scientific fields, which HHO does not use, occupy more distinct sectors: authors in probabilistic methods gather on the left of the disk, those in case-based reasoning in its upper part and those in genetic algorithms in its lower left, whereas the pairwise embedding largely mixes these fields in the lower left of the disk. This observation is qualitative, since the rule by which a field label is assigned to an author is not documented in AHORN and no field-based statistic is reported.

Of the 834 filled triangles, 706 have a defined area in both embeddings. In the other 128, as in the 1332 excluded empty triangles, two vertices coincide to numerical precision; these are almost always authors with exactly the same co-authors (158 of the 159 coincident pairs under HHO), which no pairwise or higher-order information can separate.
HHO shrinks 61.3\% of the 706 triangles, with a median area ratio of 0.577 (paired Wilcoxon $p=9.2\times10^{-18}$; figure~\ref{fig:cora}\textit{c}).
Under the pairwise embedding, the ordering of filled and empty areas is inverted: filled triangles are larger than empty ones (fold 0.236, $P=0.316$, $p=3.9\times10^{-53}$). HHO restores the expected direction, but the separation is weak (fold 1.05, $P=0.533$, $p=0.0064$; figure~\ref{fig:cora}\textit{d}).
Global distance preservation improves from $\rho=-0.0784$ [$-0.130$, $-0.0235$] to 0.417 [0.370, 0.462] (disjoint intervals, no paired test; figure~\ref{fig:cora}\textit{e}).
Group (co-author) recovery C1 increases from 0.0992 [0.0868, 0.112] to 0.163 [0.148, 0.178] ($p=2.9\times10^{-16}$, with nodes treated as independent).

\begin{figure}[!htbp]
\centering
\includegraphics[width=0.9\textwidth]{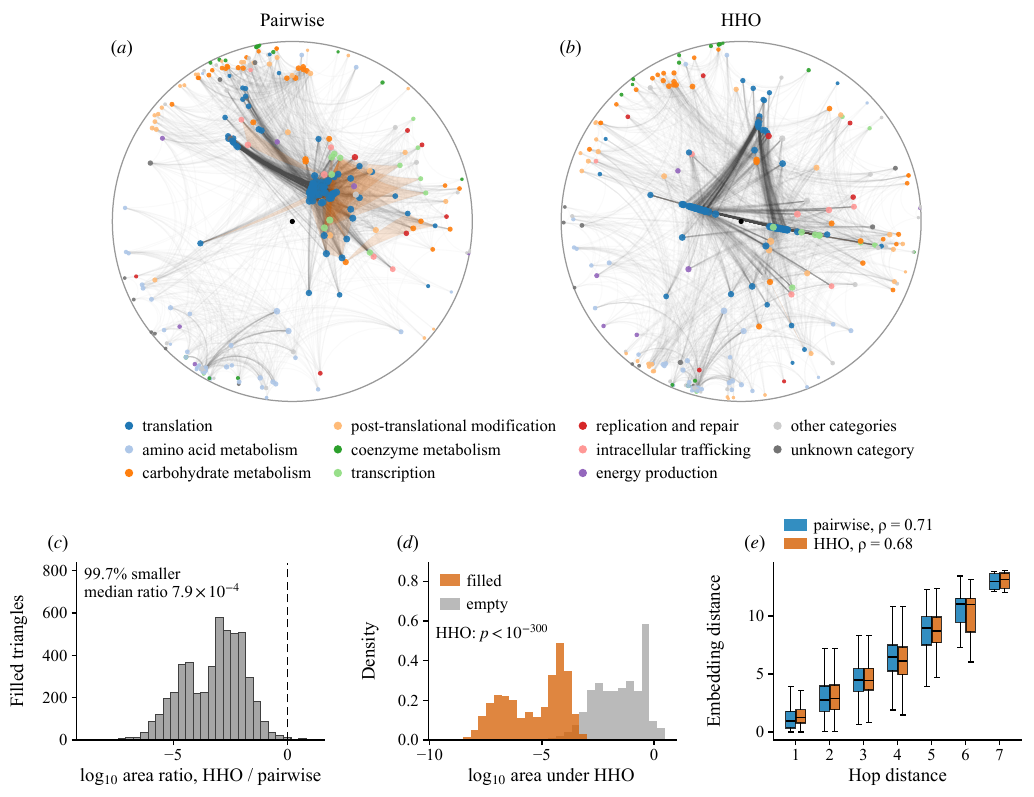}
\caption{\textbf{HHO brings members of protein complexes into compact regions of the hyperbolic disk in the protein interaction network of \textit{Escherichia coli}. 
} Giant component of 286 proteins and 5027 interactions. The simplices are the cliques formed by the members of 14 protein complexes annotated in the Complex Portal: the two ribosomal subunits, the sigma-factor variants of the RNA polymerase holoenzyme, two Sec translocon complexes, the RNA degradosome, and two transcription elongation and antitermination complexes (section~\ref{Sec:data_ecoli}). 
(\textit{a}) Pairwise and (\textit{b}) HHO embeddings, with proteins colored by COG functional category. 
Shaded regions indicate filled triangles, gray curves represent edges, and the black dot marks the disk center, as in figure~\ref{fig:synth}. The nine most frequent categories are shown in color: translation, ribosomal structure and biogenesis (84 proteins); amino acid transport and metabolism (54); carbohydrate transport and metabolism (32); post-translational modification, protein turnover, chaperones (31); coenzyme transport and metabolism (15); transcription (13); replication, recombination and repair (8); intracellular trafficking, secretion, and vesicular transport (7); energy production and conversion (6). 
The remaining nine annotated categories (25 proteins) are shown in light gray, and the 11 proteins with unknown category in dark gray.
COG categories are shown for visual reference only; no quantitative assessment of COG-category recovery was performed. (\textit{c}) 
Distribution of $\log_{10}(A_{\mathrm{HHO}}/A_{\mathrm{pairwise}})$ for the filled triangles. The dashed line at zero indicates equal areas.
Of the 5366 filled triangles, the 4941 with a non-zero HHO area are drawn; the other 425 are degenerate under HHO, with their three vertices on a common geodesic and zero area to numerical precision. 99.7\% of the 5366 are smaller under HHO. (\textit{d}) Densities of $\log_{10}$ areas of filled and empty triangles under HHO: 4941 of the 5366 filled triangles (the degenerate ones cannot be drawn on a log axis) and 69{,}346 empty triangles. $P(A_{\mathrm{filled}}<A_{\mathrm{empty}})=0.993$ under HHO and 0.704 under the pairwise embedding (not drawn); Mann--Whitney $p<10^{-300}$ for both. Many filled areas under HHO are close to zero, so only the ordering of the areas in (\textit{c}) and (\textit{d}) is interpreted (section~\ref{Sec:res_ecoli}). (\textit{e}) Embedding distance against shortest path distance (boxes as in figure~\ref{fig:synth}\textit{e}). Spearman $\rho$: pairwise 0.712, HHO 0.683.}
\label{fig:ecoli}
\end{figure}

\subsection{Protein interactions of \textit{E.~coli}: protein complexes become compact regions in the embedding space}\label{Sec:res_ecoli}
Protein complexes provide a canonical example of higher-order interactions in biological systems, as their functional activity often depends on the coordinated assembly of multiple subunits. A pairwise network representation cannot explicitly capture this collective organization, instead representing the protein complex as a set of pairwise interactions and thereby blurring its higher-order nature. On the \textit{E.~coli} protein interaction network (286 proteins and 5027 interactions), whose simplices are given by 14 protein complexes of the Complex Portal (section~\ref{Sec:data_ecoli}), HHO gathers the members of these protein complexes: 99.7\% of the 5366 filled triangles are smaller than under the pairwise embedding (figure~\ref{fig:ecoli}\textit{c}), and filled triangles are smaller than empty ones with $P(A_{\mathrm{filled}}<A_{\mathrm{empty}})=0.993$, against 0.704 for the pairwise embedding (Mann--Whitney $p<10^{-300}$ for both; figure~\ref{fig:ecoli}\textit{d}).
Under HHO, many filled triangles are compressed to near-zero areas, so that the magnitudes of the area ratio and of the fold depend on numerical resolution; we interpret only the ordering of the areas, which is preserved when these triangles are set aside ($P=0.971$).
Unlike on Holme--Kim and Cora, global distance preservation is slightly lower under HHO ($\rho=0.683$) than under the pairwise embedding ($\rho=0.712$; figure~\ref{fig:ecoli}\textit{e}; no test).

At the level of proteins, complex recovery C1 nearly doubles, from 0.374 [0.314, 0.440] under the pairwise embedding to 0.711 [0.617, 0.803] under HHO ($p=7\times10^{-10}$; 70 proteins with a partner), above Laplacian eigenmaps (0.280 [0.226, 0.343]); the hypergraph embedding of Zhou et al.\ reaches the same level as HHO (section~\ref{Sec:res_comparison}).
At the level of complexes, the mean pairwise hyperbolic distance among members decreases under HHO for all five case-study complexes (table~\ref{tab:compactness}), by a factor of 2.48 for the 50S and 2.06 for the 30S ribosomal subunits, the two largest. The five case-study complexes are the largest complexes of the giant component, taken in decreasing order of size, a complex being skipped when its Jaccard index with an already selected complex exceeds 0.3, which discards near-duplicates that share most of their subunits (such as the sigma-factor variants of the RNA polymerase).

\begin{table}[!htbp]
\caption{Compactness of the five case-study complexes of \textit{E.~coli}: mean pairwise hyperbolic distance among members under the pairwise embedding and under HHO, and their ratio (pairwise/HHO). Complexes are taken in decreasing order of size in the giant component, skipping those with a Jaccard index above 0.3 with an already selected complex.}
\label{tab:compactness}
\centering
\adjustbox{max width=\textwidth}{%
\begin{tabular}{lcccc}
\hline
complex & members & pairwise & HHO & ratio \\
\hline
50S large ribosomal subunit & 31 & 0.361 & 0.145 & 2.48 \\
30S small ribosomal subunit & 18 & 0.192 & 0.0931 & 2.06 \\
transcription elongation complex & 7 & 0.473 & 0.314 & 1.51 \\
holo-translocon SecYEG-SecDF-YajC-YidC & 3 & 0.955 & 0.558 & 1.71 \\
RNA degradosome & 3 & 1.19 & 0.921 & 1.29 \\
\hline
\end{tabular}}
\end{table}

\subsection{Comparison with other embeddings}\label{Sec:res_comparison}
We compared HHO with the four embeddings of section~\ref{Sec:method_stats}: the pairwise embedding, the weighted-clique ablation, Laplacian eigenmaps and the hypergraph embedding of Zhou et al. Table~\ref{tab:comparison} gives the value of each criterion for every method and dataset. The separation of filled and empty triangles measures most directly whether an embedding encodes the simplices. On Holme--Kim (mean over 10 graphs), HHO is the only method that separates the two classes clearly, with an AUC of 0.758 [0.739, 0.777] (higher than each of the four other methods on all 10 graphs, $p=0.002$, the smallest value attainable with 10 graphs), against 0.532 and 0.530 for the next methods, Zhou et al.\ and the weighted clique. On Cora, HHO also has the highest AUC of the five methods, 0.533 [0.509, 0.557], above the weighted clique (0.453) and the three other methods ($p=0.001$ for each), but the ratio of the median areas of empty and filled triangles (fold), 1.05 [0.893, 1.24], does not differ from 1, so that the separation remains weak, consistent with the many structurally equivalent authors. On \textit{E.~coli}, the two hyperbolic methods that take into account the simplices perform best, the weighted clique (0.998 [0.997, 0.999]) ranking ahead of HHO (0.993 [0.993, 0.994]; $p=0.001$), followed by Zhou et al.\ (0.974 [0.973, 0.975]), whereas the pairwise methods reach at most 0.704. The weighted clique and HHO both compress most filled triangles to the numerical floor, with an area below $10^{-4}$ for 99.0\% of the filled triangles under the weighted clique and 84.8\% under HHO (section~\ref{Sec:res_ecoli}), so that their ordering is not interpretable.

For group recovery C1, HHO performs better than the pairwise embedding on all three datasets (table~\ref{tab:comparison}, first two rows) and is never ranked below third of the five methods (mean rank 2.33, as Laplacian eigenmaps and Zhou et al., against 3.33 for the weighted clique). It is not the best method on this criterion: Laplacian eigenmaps improves C1 more on Holme--Kim and Cora ($+0.0798$ and $+0.214$ over the pairwise embedding), and Zhou et al.\ more on \textit{E.~coli} ($+0.349$). The weighted-clique ablation improves C1 over the pairwise embedding on Cora ($+0.0416$) and \textit{E.~coli} ($+0.300$) but not on Holme--Kim ($+0.0016$, within the confidence interval of the pairwise embedding), and HHO is above it on all three datasets (0.0788, 0.163 and 0.711 against 0.0362, 0.141 and 0.674; $p=0.002$, $0.0035$ and $0.017$). At identical radial coordinates and with the same simplices, HHO therefore recovers groups better than the addition of the simplices to the edge weights.

Several limitations qualify these results. First, on the three datasets, the filled triangles and the groups scored by the area markers and by C1 are those given to HHO, Zhou et al.\ and the weighted clique, so these criteria favor, by design, the methods that read the simplices. These criteria therefore measure how faithfully an embedding represents the groups it is given, not its ability to predict unobserved ones. Second, HHO preserves pairwise adjacency less well than some other embeddings: by edge versus non-edge AUC, it is below Laplacian eigenmaps on Holme--Kim (0.736 against 0.843) and below Laplacian eigenmaps (0.918), the weighted clique (0.891) and the pairwise embedding (0.889) on \textit{E.~coli} (HHO 0.876; $p=0.002$ on Holme--Kim and $p=0.001$ on \textit{E.~coli}). Third, on \textit{E.~coli}, the hypergraph embedding of Zhou et al.\ recovers protein complexes as well as HHO (C1 0.723 [0.626, 0.823] against 0.711 [0.617, 0.803]; $p=0.89$). Finally, the tests over nodes treat dependent nodes as independent units, which could overstate the statistical evidence.

\begin{table}[t]
\caption{
Comparison of HHO with four alternative embedding methods on three datasets. Metrics quantify the separation of filled and empty triangles (area AUC, defined as $P(A_{\mathrm{filled}}<A_{\mathrm{empty}})$, with 0.5 indicating no separation), group recovery (C1), and 
distance preservation (Spearman correlation $\rho$ between embedded distances and graph shortest-path distances). Higher values indicate better performance for all three metrics. HK denotes Holme--Kim complexes, with results averaged over 10 graphs; Ec denotes \textit{E.~coli}. For Cora, area AUC is evaluated on the common subset of triangles with a defined area (three distinct vertices) under all methods: 706 of 834 filled triangles and 3093 of 4425 empty triangles. For \textit{E.~coli}, most filled triangles have near-zero areas under both HHO and weighted clique expansion, preventing a reliable ranking of these two methods by area AUC. Boldface marks the highest numerical value in each column. The symbol $\approx$ indicates no statistically significant difference from HHO ($p\geq0.05$); all other alternative-method values differ from HHO at $p<0.05$, except the correlations $\rho$ for Cora and \textit{E.~coli}, for which no significance tests were performed. Method types specify the embedding geometry, hyperbolic (H) or Euclidean (E), and whether the input contains only the 1-skeleton (1-sk) or also higher-order simplices (HO).
}
\label{tab:comparison}
\centering
\small
\setlength{\tabcolsep}{4pt}
\adjustbox{max width=\textwidth}{%
\begin{tabular}{llccccccccc}
\hline
 & & \multicolumn{3}{c}{area AUC} & \multicolumn{3}{c}{C1} & \multicolumn{3}{c}{distance preservation $\rho$} \\
method & type & HK & Cora & Ec & HK & Cora & Ec & HK & Cora & Ec \\
\hline
Pairwise~\cite{GuillemaudMultilayer2025} & H, 1-sk & 0.508 & 0.316 & 0.704 & 0.0346 & 0.0992 & 0.374 & 0.0426 & $-0.0784$ & 0.712 \\
HHO & H, HO & \textbf{0.758} & \textbf{0.533} & 0.993 & 0.0788 & 0.163 & 0.711 & \textbf{0.391} & 0.417 & 0.683 \\
Weighted clique expansion & H, HO & 0.530 & 0.453 & \textbf{0.998} & 0.0362 & 0.141 & 0.674 & 0.0892 & 0.368 & 0.735 \\
Laplacian eigenmaps~\cite{BelkinNiyogi2003} & E, 1-sk & 0.489 & 0.393 & 0.546 & \textbf{0.114} & \textbf{0.313} & 0.280 & 0.356 & \textbf{0.435} & \textbf{0.770} \\
Hypergraph spectral~\cite{Zhou2006} & E, HO & 0.532 & 0.388 & 0.974 & 0.108 & 0.134 & $\approx$\textbf{0.723} & 0.204 & 0.0142 & 0.630 \\
\hline
\end{tabular}}
\end{table}

\section{Discussion}\label{Sec:discussion}
We have introduced HHO, a method 
for embedding simplicial complexes 
in the hyperbolic disk 
through a two-layer network representation. The first layer of the network contains the original nodes and their pairwise links; the second represents the filled triangles belonging to the simplicial complex. 
Because the radial coordinate of each node is
determined  
by its strength in the 1-skeleton, the higher-order information 
shapes its angular position only. The hyperbolic area of a triangle then quantifies 
how compactly a group interaction is represented.
Across all four complexes, HHO 
reduces the area of most filled triangles relative to the pairwise embedding. 
Filled triangles become clearly smaller than empty ones on both synthetic complexes and on the \textit{E.~coli} protein interaction network, whereas the separation remains weak on the Cora co-authorship complex. For the group recovery, HHO is competitive with, 
but does not consistently outperform, 
Euclidean methods based on 
the 1-skeleton or the hypergraph. Nevertheless, HHO recovers groups better than the weighted-clique ablation on all three datasets, in which the same simplices are incorporated into the edge weights of the pairwise embedding while the radial coordinates are kept unchanged. This result indicates that the observed improvement cannot be attributed solely to the additional weighting of pairwise connections, but rather highlights the contribution of the two-layer network construction, which explicitly represents higher-order interactions.

The main contribution of HHO is therefore a representation of higher-order structure in a geometric space whose coordinates have a known interpretation. In hyperbolic maps of networks, the radial coordinate reflects the popularity or hierarchical level of a node and the angular coordinate its similarity to other nodes~\cite{Krioukov2010,Papadopoulos2012,Bogu2021}. HHO preserves this reading and adds to it a property of groups: the members of a group interaction are drawn towards a common angular sector, at the radial positions given by their connectivity. Since the output of HHO is a set of coordinates in the Poincar\'e disk, it can be analyzed directly with the tools developed for hyperbolic network maps, such as angular community detection~\cite{Faqeeh2018}, greedy routing~\cite{Bogu2010,Allard2020}, link prediction~\cite{Kitsak2020} or geometric renormalization~\cite{GarciaPerez2018}. In brain networks, hyperbolic coordinates have been used to detect local perturbations and regions disrupted by neurodegeneration~\cite{Longhena2024,Longhena2025}, and to forecast epileptic seizures and the outcome of epilepsy surgery~\cite{GuillemaudSeizure2025,GuillemaudEpilepsy2025}. The second layer of HHO is related to the bipartite incidence graph of group interactions, whose latent geometry has been studied on its own~\cite{Kitsak2017}; HHO couples this incidence structure to the pairwise network instead of embedding it separately. Unlike hyperbolic analyses of multiplex networks, in which each layer is embedded separately~\cite{Kleineberg2016}, HHO embeds its two layers jointly.

The \textit{E.~coli} case study illustrates this reading on a biological system. Protein complexes appear as compact angular sectors of the map: the 50S and 30S ribosomal subunits, for instance, are about 2.5 and 2.1 times more compact under HHO than under the pairwise embedding, while the radial coordinate still orders the proteins by connectivity. This organization recalls the modular and economical organization of biological networks, in which densely interacting units are grouped into modules at a low wiring cost\cite{Bullmore2012}. This parallel is an analogy only: an area in the latent space is not a physical wiring length, and the case study, restricted to a subnetwork dominated by the translation and transcription machinery, is intended to illustrate the method rather than to characterize the \textit{E.~coli} complexome.

Several aspects of the method  
warrant further investigation.  
First, the pairwise embedding used as reference combines a coalescent-type angular embedding~\cite{Muscoloni2017} with our radial rule, and its distances 
correlate poorly with 
shortest path distances on the Holme--Kim and Cora complexes ($\rho=0.0426$ and $-0.0784$; table~\ref{tab:comparison}). The gains of HHO should therefore also be assessed against other hyperbolic embedding frameworks, such as Mercator~\cite{Garca-Prez2019}. Second, on \textit{E.~coli}, many filled triangles are compressed to near-zero areas, so that only the ordering of the areas, and not their magnitude, is interpretable. Third, the two-layer network representation uses a dense $(N+M)\times(N+M)$ matrix, which restricted the \textit{E.~coli} analysis to 300 proteins. Sparse matrices and landmark-based variants of Isomap would allow larger simplicial complexes to be embedded. Finally, the parameters $\mu$ and $\gamma$, which set the weight of the second layer, were fixed per dataset and not optimized.

In its present form, HHO incorporates interactions among more than three nodes, i.e.\ simplices of dimension $k\geq3$, only through their triangular faces. It therefore does not distinguish simplicial complexes that share the same 2-skeleton (the same nodes, edges and filled triangles) but differ in their higher-dimensional simplices. A possible extension would introduce a separate layer for each simplex dimension, explicitly representing interactions beyond filled triangles, with the hyperbolic volumes of $k$-simplices, $k\geq3$, as markers.
Neuroscience offers a promising application, as 
  cliques and cavities of structural and functional brain networks carry information that is invisible at the level of single edges\cite{Giusti2016,Petri2014,Sizemore2018,Reimann2017}, and hyperbolic maps of brain networks already relate their geometry to anatomy and pathology\cite{Allard2020,Cacciola2017,Longhena2025,GuillemaudEpilepsy2025}. 
A further extension could concern temporal simplicial complexes, where group interactions appear and disappear, and the trajectory of each group in the hyperbolic disk could be followed over time.

\section*{Data accessibility}
The HHO method is available in the open-source Python package HOHypEmb, version 0.1.0 (\url{https://github.com/MartinG-38/HOHypEmb}; archived on Zenodo\cite{HOHypEmb2026}; BSD-3-Clause license), together with example notebooks that compute the embeddings of the four complexes with the parameters used in this article. The Cora co-authorship complex is available from the AHORN repository\cite{Frantzen2026,AHORNCora2026}, protein associations from STRING v12.0\cite{Szklarczyk2023} (\url{https://string-db.org}, taxon 511145), protein complexes from the Complex Portal\cite{Meldal2022} (\url{https://www.ebi.ac.uk/complexportal}, taxon 83333) and annotations from the COG database\cite{Galperin2021}.

\section*{Declaration of AI use}
Generative AI tools were used to assist with code development and with the review of the manuscript; the authors checked all outputs.

\section*{Authors' contributions}
M.G.: conceptualization, methodology, software, formal analysis, writing; M.C.: conceptualization, methodology, supervision, writing; V.L.: supervision, writing. All authors gave final approval for publication and agreed to be held accountable for the work performed therein.

\section*{Funding}
M.G. was supported by a doctoral fellowship from the \'Ecole Normale Sup\'erieure de Lyon and Inria.

\bibliographystyle{unsrt}
\bibliography{biblio}

\end{document}